\documentclass[aps,prd,twocolumn,groupedaddress,eqsecnum,amssymb,showpacs,
nofootinbib]{revtex4}
\usepackage{graphicx,mathptm,bm,times,epsfig}

\newcommand{\bea}{\begin{eqnarray}}
\newcommand{\eea}{\end{eqnarray}}

\begin{document}

\title{Probing Dark Energy with Black Hole Binaries}

\author{Laura Mersini-Houghton and Adam Kelleher}
\email[]{mersini@physics.unc.edu, akelleh@physics.unc.edu}
\affiliation{Department of Physics and Astronomy, UNC-Chapel Hill,
NC, 27599-3255, USA}

\date{\today}
 
\begin{abstract}

The equation of state (EoS) of dark energy $w$ remains elusive despite enormous experimental efforts to pin down its value and its time variation. Yet it is the single most important handle we have in our understanding of one of the most mysterious puzzle in nature, dark energy. This letter proposes a new method for measuring the EoS of dark energy by using the gravitational waves (GW) of black hole binaries. The method described here offers an alternative to the standard way of large scale surveys. 

It is well known that the mass of a black hole changes due to the accretion of dark energy but at an extremely slow rate. However, a binary of supermassive black holes (SBH) radiates gravitational waves with a power proportional to the masses of these accreting stars and thereby carries information on dark energy. These waves can propagate through the vastness of structure in the universe unimpeded. The orbital changes of the binary, induced by the energy loss from gravitational radiation, receive a large contribution from dark energy accretion. This contribution is directly proportional to $(1+w)$ and is dominant for SBH binaries with separation $R \ge 1000$ parsec, thereby accelerating the merging process for $w > -1$ or ripping the stars apart for phantom dark energy with $w < -1$. Such orbital changes, therefore $w$, can be detected with LIGO and LISA near merging time, or with X-ray and radio measurements of Chandra and VLBA experiments. 

\end{abstract}

\pacs{98.80.Qc, 11.25.Wx}

\maketitle

\section{Introduction} 
\label{sec:intro}

One of our most crucial questions about nature at present is: what is dark energy? The fact that our universe is accelerating \cite{wmap,lauramelchiorri} and that dark energy constitutes about $70 percent$ of the total energy density of the universe, are well established by now. Many theoretical models have been put forth which cast dark energy in the form of a cosmic fluid \cite{quintessence,kessence,transplanck} with time variations in its equation of state $w(z)$. Yet the simplest explanation for dark energy remains to be a pure cosmological constant (cc) $\Lambda$. 

The trouble we face in understanding dark energy does not stem from a shortage of dark energy models, with $w(z_0) \simeq -1$, that mimick at present the behavour of $\Lambda$ and give rise to the observed acceleration of the universe. The puzzle rather lies on identifying which one of these possible candidates is the correct one. The best way to discriminate among the various possibilites and a pure $cc$, $\Lambda$, is to experimentally measure the time variations of the dark energy equation of state $w(z)$. So far a popular parametrization for $w(z)$ is the linear one $w(z)=w_0 + w_{1} z +...$ with $w_1 = 0$ for a pure {\it cc} \cite{eos}. A knowledge of $w(z)$ is crucial for not only understanding the present accelerated expansion of the universe but also for making predictions for its future evolution and destiny. If $w(z) \ll -1$ then dark energy is a phantom \cite{phantom} which leads the universe to a Big Rip in the future. If $w = -1$ then we are probably \cite{laurads} facing an eternal DeSitter state \cite{eternalds} which at least at the classical level implies constant temperature and entropy therefore a cosmic heat death \cite{fredlaura, fred}. Other forms of $w(z)$ can also allow for a Big Crunch \cite{crunch} or bounces \cite{bounce}. At present we can not infer which destiny our universe will meet without a better knowldege of $w(z)$. 

Major experimental efforts for pinning down the value and time-variations of $w(z)$ are under way through large scale surveys from CMB \cite{wmap}, large scale structure \cite{SDSS}, and SN1a observations. The endeavor of measuring, to confident precision, such small time variations in $w(z)$ has proven extremely difficult, partly due to the inherited errors in the experiments that are not instrumental but which originate from noise accumulated from the background and foreground effects through which the signal we receive has propagated. In order to minimize such errors associated with the propagation of the signal through the vastness of structure in the universe, we would like to propose in this letter a complimentary method for measuring $w(z)$. This method uses compact and localized objects, such as Black Holes, for acquiring information about $w(z)$,  by exploiting the gravitational waves these objects emit when they are in binaries. The advantage of this method is twofold: first, gravitational waves propagate undisturbed through structure; and second, we have existing experiments which are either already operational or will be in the near future, such as LIGO and LISA missions designed to detect these binaries gravitational waves, or Chandra and VLBA experiments designed for X-ray and radio measurements.

The accretion of dark energy by Black Holes is reviewed in Sec.2., including a review of the main parameters of binaries and gravitational waves, useful for our purposed. Sec.3presents the method we propose, along with an investigation, discussion and some illustrations, on how gravitational waves from SBH's binaries can be used for extrapolating the equation of state $w(z)$ of dark energy.

\section{Black Hole Binaries}

The phenomenon of accretion of dark energy by black holes is now established. If dark energy is modeled as a background cosmic fluid, then the flux of energy accreted by the hole will change the mass of the black hole $M$ by a rate given by the equation \cite{babichev, magueijo}


\begin{equation} 
\dot{M} = 4 \pi A M^2 \rho_{\infty}
\left[ 1+ w(z) \right] \, , 
\label{bhmass}
\end{equation}
where $A \simeq 5.6$ is a numerical factor whose value depends on the matching done for the dark energy fluid velocity at the horizon of the black hole, $\rho_{\infty}$ is the energy density of dark energy far away from the hole, and $w(z)$ is the equation of state for dark energy as a function of the redshift $z$.
For the case of a quintessence field, Eqn.\ref{bhmass} becomes $\dot{M} = \pm 4\pi (2M)^2 \dot{\phi}^2$ where the $(-)$ sign corresponds to phantom fields\cite{babichev}. Clearly by accreting dark energy black holes gain in mass for $w > -1$ and lose mass for $w < -1$. The solution to Eqn.\ref{bhmass} is
\begin{equation}
M ={ M_0 \over \left[1-\frac{t}{\tau} \right] } \ ,
\label{bhsol}
\end{equation}
where $M_0$ is the initial mass of the black hole and $\tau$ is its evolution time scale

\begin{equation}
\tau = \frac{1}{4\pi A M_{0}\rho_{\infty}(1+w)} \ ,
\label{tau}
\end{equation}

In the case of phantom energy, $\tau$ provides the time scale to the Big Rip, by which time all black holes must have completely evaporated.

If dark energy is a quintessence scalar field with $w>-1$, the authors of \cite{magueijo} have shown that the growth of the mass of the black hole due to accretion of dark energy could provide a sufficient mechanism for converting primordial black holes (PBH) into supermassive black holes (SBH) within a reasonable time, a time comparable to the age of the universe. SBH's  can be produced via dark matter accretion \cite{ostriker} and they can have a mass as large as $10^{8-18} M_{\ast}$ where $M_{\ast}$ is a solar mass.

\subsection{SBH Binaries}

Most active galactic nuclei (AGN) are populated by SBH's therefore we can expect that SBH binaries are not that uncommon. A binary of two massive objects will emit graviational 
waves with angular frequency $\Omega$
\begin{equation}
\Omega = 2 \left[\frac{G(m_1 + m_2}{R^3} \right]^{1/2} \ ,
\label{omega}
\end{equation}
where $R$ is the separation of the stars in the orbit, $G$ is Newton's constant, and $(m_1 , m_2)$ are the masses of the stars in the binary. 
The period of rotation is $T = {2\pi \over \Omega}$. Another useful detection parameter is the amplitude $h_+ = \frac{G m_{1}m_{2}}{r R}$ where $r$ is the distance along the line of sight to the observer.

The power $P_{GW}$ contained in the gravitational waves results in a loss of the energy of the binary $E=Mc^2$ \cite{thorn} where

\begin{equation}
P_{GW} = \frac{-32 G^4}{5 c^5}\left[\frac{m^{2}_{1}m^{2}_{2}(m_1 +m_2)}{R^5} \right] \ ,
\label{power}
\end{equation}

and the effective gravitational mass of the binary $M$ is defined by
\begin{equation}
M = m_{1} + m_{2} -\frac{1}{2} \frac{m_{1} m_{2}}{R} \ ,.
\label{effectivemass}
\end{equation}

As a result the orbits get smaller until finally the merging occurs. {\it LIGO} and {\it LISA} experiments are designed to look for such gravitational waves, while {\it VLBA} and {\it Chandra} experiments can observe the binaries via radio and X-ray measurements. The optimal {\it LIGO} band for detection is in frequencies around $\Omega = 150 Hz$  \cite{ligo} and {\it LISA} will be able to detect waves with frequencies as low as $10^{-6} Hz$.

\section{Information on Dark Energy from Black Hole Binaries}

In order to illustrate the main idea of the method proposed here, let us for simplicity take equal mass binaries, $m_{1} = m_{2} = m$ of supermassive black holes that are in the background of the unknown dark energy fluid. In this case the effective mass from Eqn. \ref{effectivemass} is: $M = 2 m - \frac{m^2}{2 R}$, and the expression Eqn.\ref{power} for the power of the emitted gravitational waves (GW), becomes

\begin{equation}
P_{GW} = \frac{-32}{5} \left[\frac{2 m^5}{R^5} \right] \ ,
\label{powersimple}
 \end{equation}

Since the binary is immersed in the 'bath' of dark energy with energy density $\rho_{\Lambda}$, the mass of each hole in the binary will change according to Eqns.\ref{bhmass}, \ref{bhsol}, due to the accretion of dark energy             

\begin{equation}
\dot{m} = m^2 4\pi A \rho_{\Lambda} (1+w)
\equiv 2 m^{2} L \ ,
\label{masssimple}
\end{equation}
Here $w(z)$ denotes the equation of state of dark energy at redshift $z$, and $m_0$ the mass of the star at initial time $t=0$. The parameter $L$ is defined such that: $2 L m_0 (\frac{G^2}{c^3}) = \tau^{-1}$, with the evolution time $\tau$ given in Eqn. \ref{tau}. If the mass of black holes in the binary changes then the power of the emitted gravitational waves $P_{GW}$ has to change accordingly, due to the accretion of dark energy. But the energy losses from $P_{GW}$ prior to the inclusion of dark energy accretion and after, result in quite different orbital change and in some cases in a highly different merging timescale, (see Fig.1 and Fig.2). These orbital changes due to dark energy depend sensitively on $\rho_{\Lambda} , w(z), m_0$

We can estimate the dependence of $R$ as a function of $w$ and its time evolution, $R[t, w]$ by solving a differential equation which is derived by matching the mass changes $[ dMc^2 / dt]$ which now are due to both: dark energy and GW's, with the energy losses $P_{GW}$ through gravitational waves, 

\begin{equation}
P_{GW} = \frac{d(Mc^2)}{dt} = \frac{-32}{5} \left[\frac{2m^5}{R^5}\right] \ ,
\label{matchpower}
\end{equation}

The effect of dark energy accretion on the mass of stars Eqn. \ref{masssimple} is now included in the expression for power Eqn\ref{powersimple} and the matching in Eqn. \ref{matchpower} results in the following differential equation

\begin{eqnarray}
& & R^3 \frac{dR}{dt} = {-64 \over 5} \frac{2 m_{0}^3}{ [1- 2 L t m_0]^3 }  \nonumber \\
&- & \left[\frac{- 4 L  R^4 m_{0} }{[1 - 2 L t m_{0}] }  + 8 L R^6  \right ] \ ,
\label{diffeqn}
\end{eqnarray}

This equation, along with its solution below, Eqn. \ref{neworbit2}, are the most important results of this investigation.
Notice that the first term for $m=m_{0}$ corresponds to the well known general relativity effect of gravitational waves emission, the Hulse-Taylor effect. But the other two terms are new and completely due to the dark energy accretion by black holes. 

Clearly the orbital separation $R[t,w]$ depends sensitively on the dark energy equation of state $w(z)$. We can thus use observations of the orbital changes $R$ with time in order to extract information  about dark energy's EoS $w(z)$ by using the already existing and planned GW's experiments, {\it LIGO} and {\it LISA}, or directly via $X-ray$ measurement missions like Chandra and radio measurements with {\it VLBA} \cite{vlba}.

The changes in the frequency $\dot{\Omega}$  and therefore the orbital period $T = \frac{2\pi}{\Omega}$ can be derived from Eqn.\ref{matchpower} in the same manner, which yield

\begin{equation}
\Omega = \Omega_0 (\frac{R_0}{R[t,w]})^{3/2} \ ,
\end{equation}

where $R_0$ is the initial separation and $\Omega_0$ initial angular frequency at $t=0$. If the orbits get smaller with time, $R[t,w] \ll R_0$, then according to Eqn.3.5. the frequency $\Omega$
might increase sufficiently such as to fall within the detection limits of {\it LIGO} or {\it LISA}. From solutions to Eqn.\ref{diffeqn} we can get an estimate for the effect of dark energy on parameters of the SBH's binary.

To a first order approximation the solutions of Eqn.\ref{diffeqn} is
\begin{eqnarray}
& & R[t,w] = [R_0^{4} + 8 R_0^{4} {\rm Log}(1-2 L t m_0) - 32 L R_0^{5} t \nonumber \\
&-& \frac{32}{5}({16 \over L}) \left[\frac{m_0^{2}}{(1 - 2 L t m_{0})^2} - m_0^{2} \right] ]^{1/4} \ ,
\label{neworbit1}
 \end{eqnarray}

In general $\tau$ is quite large. For example, for a solar mass $M_{\ast}, \tau \simeq 10^{32} yrs$. Therefore to a good approximation, we can re-write Eq.\ref{neworbit1}, including the appropriate units, as

\begin{eqnarray}
& & R[t,w] = R_{0} [1 + 16 L m_{0}(\frac{G^2}{c^3}) t - \nonumber \\
& -&  32 L R_{0}(\frac{G}{c}) t - \frac{64}{5}({4 G^{3}\over c^5})\left[\frac{8 m_0^{3}}{R_{0}^4} t \right ] ]^{1/4} 
\label{neworbit2}
 \end{eqnarray}

which for ${t \over \tau} = 2 L t m_0 \ll 1$ recovers the Hulse-Taylor corrections to orbital changes given by the last term, that are linear in time $t$. In the limit of low mass stars,  the dark energy correction terms in Eqn. \ref{neworbit2}, are subdominant.

\begin{eqnarray}
R[t,w] = R_{0} \left[1 + 8 \frac{t}{\tau} - 16 \frac{R_{0} t}{m_{0}\tau} - \frac{64}{5}\frac{16 m_{0}^3}{R_{0}^4} t  \right]^{1/4}  \ ,
\label{neworbit3}
 \end{eqnarray}

For this reason, we focus on the large mass limit below and show that the dark energy correction terms to the orbit, Eqn. \ref{neworbit2}, are significant and even dominant for SBH's binaries with $R_0 \simeq 1000 pc , a \simeq 10^8$ or larger.

All the information about dark energy is contained in $L ={1 \over 2\tau m_0} = 2\pi \rho_{\Lambda}(1+w)$ which is positive for all dark energy models with $w \ge -1$. For the case of a pure cosmological constant (cc), $L \equiv 0$, and the corrections due to dark energy disappear entirely. However for phantom dark energy where $w < -1$, $L$ is negative. The correction term due to dark energy in Eqn.\ref{neworbit1}, in the case of phantom dark energy, thus has the opposite sign to the third (Hulse-Taylor) term that originates from power losses due to emission of gravitational waves. The sensitivity of the binary to the dark energy EoS $w(z)$ is now clear: orbital changes of a binary immersed in phantom energy $L <0$ increase the separation and, are quite different from the orbital changes in the case of dark energy with $L>0$ which accelerate the merging, or the case $L\equiv 0$ where no changes of the orbit are induced from dark energy.
As shown below, this result is used to deduce whether dark energy is a cc, a phantom or whether $w$ lies above the cc boundary $w\equiv -1$.

In order to get a family of examples, let us parametrize the masses of SBH's and the orbital separation by the following relations

\begin{equation}
m_0 = a M_{\ast} \approx a 10^{30} kg
\label{mass}
\end{equation}

where the parameter $a$ quantifies how heavy the SBH is relative to the solar mass $M_{\ast}$, and

\begin{equation}
R_0 = 2 m_0 \beta ({G \over c^2}) \approx 2\beta a 10^3
\label{orbitmass}
\end{equation}

where the parameter $\beta \gg 2$ reinforces the requirement that the orbital separation better be larger than the Schwarzchild radius $2m_0$ of each star at initial time $t=0$, long before merging.

Since the evaporation time scale $\tau$ for a solar mass black hole ($a=1$) is about $\tau = 10^{40} s = 10^{32} yrs $, then the lifetime $\tau$ for the supermassive black holes with mass given by Eq.\ref{mass} is
 
\begin{equation}
\tau = {10^{40} \over a} s = {10^{32}\over a} yrs
\end{equation} 

while the frequency of the emitted GW's, $ f={\Omega \over 2\pi} $, and the amplitude $h$ in terms of these parameters $a,\beta$ become

\begin{equation}
f_0 = \frac{10^5}{(2\beta)^{3/2} a}
\label{frequency}
\end{equation}

and

\begin{equation}
h = \frac{1}{r} {2\over \beta} a 10^3
\label{h}
\end{equation} 

LIGO is designed to detect signals in the range $f = (100, 1000) Hz$ and amplitudes $h$ around $10^{-23}, 10^{-26} m^{-1}$, with its optimum detection at frequencies $f \approx 150 Hz$ \cite{ligo}, while LISA will be able to see as far as $f = 10^{-6} Hz$. One such example of an optimum signal for LIGO \cite{ligo} would be a black hole binary with parameters: $a=10, \beta=10^{4/3}, r = 125 Mpc$. Although most of the SBH's binaries, in which the dark energy correction term $y_2$ is dominant, do not initially fall within the frequency detection limits of {\it LIGO}, they can still be detected during the time close to merging, since their frequency evolves as

\begin{equation}
\Omega = \Omega_{0} (\frac{R[t,w]}{R_0})^{3/2} 
\label{omegalater}
\end{equation}

If, for example, we estimate the orbital changes during a time $t \simeq H_0^{-1}$ with $H_0$ the Hubble constant due to both correction terms, then
the orbital separation of a binary with initial frequency $\Omega_0 = 10^{-16} Hz$, $R_0 = 5000 pc , a = 10^{12}$ will lately be $\Omega = 10^{-2} Hz$, i.e. it will fall within the detection limits.

It can be shown that the first correction term in Eq.\ref{neworbit2} is roughly ${1 \over \ (2\beta)}$ times the second correction term, therefore small enough to the second correction term that it can safely be neglected.

The interesting part in Eq.\ref{neworbit2} is the comparison between the second correction term $y_2$ which is the new term derived in this letter and is due to the dark energy 'bath' ,and the third correction term $y_1$ that has already been known and is solely due to the power losses from GW emission. Hereon, we include only the second term, $y_2$, referred to as the dark energy term, and the third term, $y_1$, referred to as the GW (Hulse-Taylor) term.

Replacing the values for the approriate factors in Eq.\ref{neworbit2} we obtain

\begin{equation}
y_{1} = - \frac{10^7}{(2\beta)^4 a} t 
\label{y1}
\end{equation}

\begin{equation}
y_{2} = - 10^{-38} (1+w) (2\beta) a t \simeq \pm 10^{-39} 2\beta a t 
\label{y2}
\end{equation}

where in the last step, the dark energy equation of state is taken to be $(1+w)\simeq \pm 0.1$ with the sign flipped for phantom dark energy, $(1+w) < -0$.

For comparison, the ratio of the two correction terms is

\begin{equation}
{y_{1} \over y_{2}} \simeq \frac{10^{45}}{(2\beta)^5 (1+w) a^2}
\label{ratio}
\end{equation}

From Eqs.\ref{y1}-\ref{ratio}, we can see that the correction term due to dark energy $y_2$ can be as large as the previously known GW (Hulse-Taylor) correction term $y_1$, or even dominant in some cases, for supermassive black holes $a\gg 1$ and for large orbital separation $\beta \gg 2$.

Making use of this parametrization, Eqns. \ref{y1}-\ref{ratio}, we arrive at an important conclusion: all binaries that satisfy the condition
\begin{equation}
2 \beta a \ge 10^8
\label{condition}
\end{equation}

would have merged if dark energy lies above the cc boundary, $(1+w) >0$; or split apart for phantom dark energy $(1+w) < 0$. From those, from Eqn. \ref{ratio}, all binaries with separation $2 \beta \ge 10^3$ which corresponds to an initial separation $R_0 \ge 5000 pc$ are dominated by the dark energy correction term, $y_2$. {\it So, Eqn. \ref{condition} seems to provide a cutoff for SBH's binaries, which heavily depends on the type of dark energy and whether dark energy lies below or above the cc-boundary of $w=-1$}.
In general, measurements of $R[t,w]$ can reveal the equation of state of dark energy $(1+w)$, via the Eqns. \ref{y2}-\ref{ratio}, and binaries that fall under the category of Eqn. \ref{condition}, can immediately reveal whether dark energy is a phantom or not. For the other SBH's binaries, this information is deduced from

\begin{equation}
1 - \left(\frac{R[t,w]}{R_0} \right)^4 = [ y_1 + y_2]
\label{orbitchange}
\end{equation}
Considering that the total mass in the universe is $10^{55} kg$, then the rough number of SBH's is around $10^9$ or one SBH for $10^{-3} Mpc$. If we assume that most of them are in binaries, then Eqn. \ref{condition} implies that if dark energy is a phantom then we should find twice as many SBH's with separation larger than $5000 pc$ as compared to the number of SBH's we would find if dark energy is $(1+w) > 0$ for which case many of them would have merged to create even heavier SBH's.

Let us take some specific examples to illustrate this effect:

For a binary with $2\beta = 10^8, a = 10^4$ that correspond to an orbit $R_0 =10^{15}m \approx 0.1 pc$ and period of rotation $T = 1/f = 10^3 yrs$, the GW correction term is of the order $y_1 = 10^{-33} t$ while the dark energy correction term is $y_2 = 10^{-27} t$. In this example the correction term due to dark energy is many orders of magnitude larger than the GW correction and results in a net orbital change of $\delta R \simeq 10^8$ during a Hubble time.
It is important to point out in this example that if dark energy is a phantom then $y_2$ has a positive sign, while $y_1 < 0$. Since $y_2 \gg y_1$ then the merging of the two stars in the binary would not occur, due to the effect of the phantom energy dominating over and compensating for the GW power losses. In fact, since $R[t,w]$ increases in this case Eq.\ref{neworbit2}, the stars in the binary would be 'ripped apart' over time. The lifetime of the stars $\tau$ is equal to the Big Rip time in the phantom energy case and therefore larger then the characteristic times of the system, $T$ and,  it is larger than $t_\ast$, where $t_\ast$ is defined as the time when $R[t_\ast , w] \ge R_0$. But, if dark energy is not a phantom or a cc $\Lambda$, then the dark energy corrections accelerate the merging process and for $2\beta a \ge 10^{18}$ dominate it, since in this case $y_2 > y_1$. By measuring the change in the orbital separation, using GW observations with {\it LIGO} and {\it LISA} or radio measurements with {\it VLBA}, we can deduce whether $(1+w)$ is positive or negative, since for this class of SBH's binaries, the dark energy corection term $y_2 > y_1$ dominates the orbital changes.

However, the frequency $f_0 = 10^{-11} Hz $ at initial times $t=0$ of this binary is not within the LIGO or LISA limits of detection. Yet, we can use $X-ray$ or radio frequency measurements with Chandra \cite{chandra} to detect changes in the phases of signals emitted from the SBH's binary.

\begin{figure}[!htbp]
\begin{center}
\raggedleft \centerline{\epsfxsize=3.9in \epsfbox{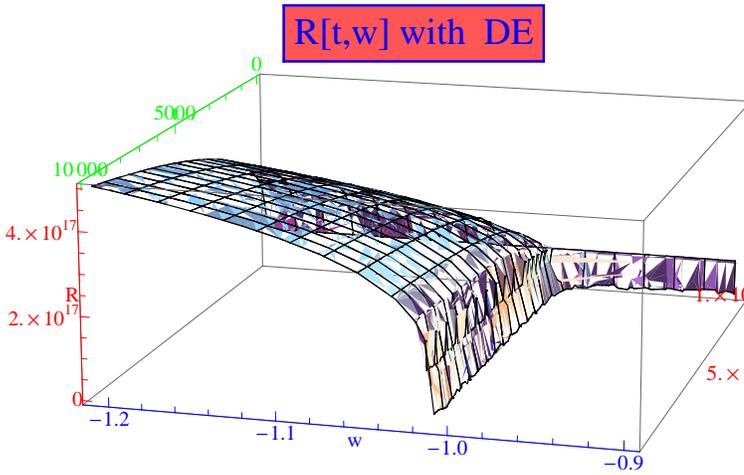}}
\caption{ Orbital separation of the SBH's binary $R[t,w]$ as a function of time and of the dark energy's $w$. The parameters are chosen to be those of the {\it Galaxy 0402+379} binary, with initial separation $R_0 \simeq 10^{17} m$. The range of $w$ shown in the plot is taken: $-0.14 \le [1+w] \le 0.14$. This figure shows the evolution of the binary's separation with time, for the case when the new corrections due to dark energy $y_2$ derived here, are included.} 
\label{fig:fig1}
\end{center}
\end{figure}

\begin{figure}[!htbp]
\begin{center}
\raggedleft \centerline{\epsfxsize=4.0in \epsfbox{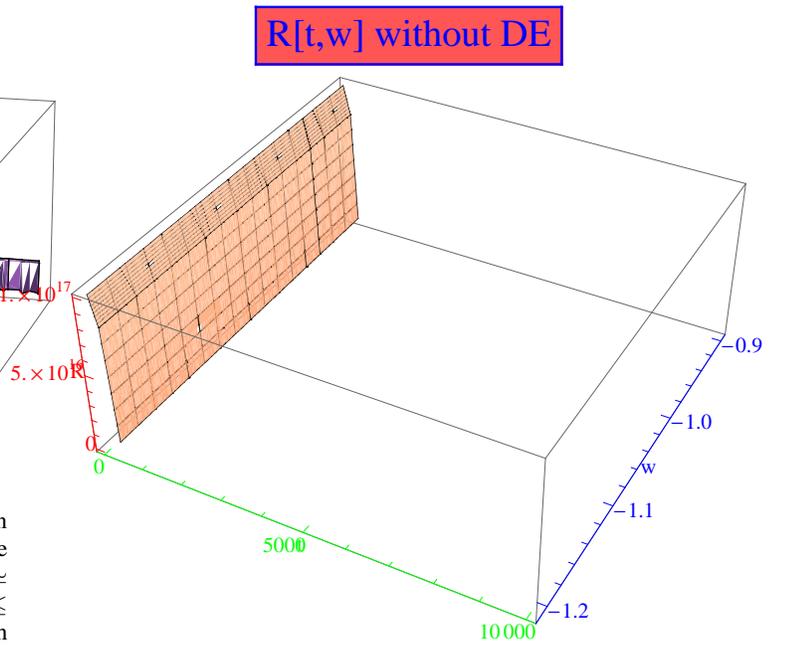}}
\caption{Orbital separation of the SBH's binary $R[t,w]$ as a function of time and of the dark energy's $w$. The parameters are chosen to be those of the {\it Galaxy 0402+379} binary, with initial separation $R_0 \simeq 10^{17} m$. The range of $w$ shown in the plot is taken: $-0.14 \le [1+w] \le 0.14$. This panel plots the same binary as the one in Fig.1 but without the corrections induced by dark energy. The difference between the two plots is clear: for $(1+w) >0$ the merging occurs faster in the ``top'' plot due to effects of $y_2$; for the phantom case $(1+w)<0$, the corrections $y_2$ slow down and even reverse the merging that the GW effects of the $y_1$ term are trying to induce.} 
\label{fig:fig2}
\end{center}
\end{figure}

Let us look at a more extreme class of binaries that satisfy: $2\beta a \approx 10^{18}$. We have $R_0 \simeq 5000 pc$, $y_2 \simeq 10^{-20} (1+w)t, y_1 \simeq 10^{-29}/(2\beta)^3 t, z \simeq \pm 10^{-8}$ for $(1+w) \simeq \pm 0.1$. The initial frequency is $\Omega_0 \simeq 10^{-16} Hz$ and for non-phantom dark energy $(1+w) > 0$, during $t= H_{0}^{-1} \simeq 10^{-20} s$ we have $R \simeq 10^{-10} R_0$, therefore $\Omega \simeq 10^{-2} Hz$ falls within current detection limit. If dark energy is a phantom the orbit would be ripped apart by that amount, then the frequency would be extremely small and with no chance of detection.

SBH's binaries have already been observed. We now describe two of them which have been observed in the last 2 years:
 
i) The first example is:{\it GALAXY 0402+379} observed in 2007 with {\it VLBA} which has the following parameters, $R_0 = 10^{17}m, T \simeq 10^{14} s, 2\beta =10^6, a = 2 10^8, r = 10^{26}m$. This binary thus has a frequency $\Omega_0 \simeq 10^{-14} Hz$ too small for LIGO detection and amplitude $h \simeq 10^{-23}$. The orbital corrections due to GW radiation $y_1$ and dark energy accretion $y_2$ differ by two orders of magnitude, $y_1 \simeq \pm {y_2 \over 10^2} \simeq 10^{-24.5} t$ resulting in an orbital change of $\delta R \simeq (1+w) 10^6 m$ during each period of rotation, $t = T$. Since the time the signal was emitted from the binary, $t_e = {r \over c} \simeq 10^{14} s$, the orbit has changed by $10^{6.5} m$ due to the GW (Hulse-Taylor) term $y_1$, and by $10^8 m$ due to the dark energy accretion term $y_2$. Thus the merging time for this binary, if the dark energy effect is ignored, is about another $60,000 yrs$ since the time the emitted signal $t_e$ was received. Included the effect of dark energy accretion results in a merging time $1,000 yrs$ for $(1+w) >0$, which is roughly two orders of magnitude less then the standard estimate. If $(1+w) <0$ then merging can not occur since $y_2$ splitting the stars apart for this case, is two orders of magnitude larger than $y_1$ which is trying to induce merging. The evolution lifetime of each star in this binary is many orders of magnitude larger than the characteristic merging scale, $\tau \simeq {10^{32} \over a} yrs \simeq 10^{24} yrs$. The latter illustrates the fact that while mass changes due to dark energy accretion by individual SBH's are too small to give rise to detection, a binary of SBH's can yield information on dark energy which is easily accessible by our current and near future experiments.

It is quite amazing that we can obtain such a wealth of information on the mysterious dark energy in the universe by using local objects such as SBH's binaries, and experiments that were not initially designed for dark energy detection, such as {\it LIGO, LISA, VLBA and Chandra} experiments. This binary's separation as a function of time and $w$ is plotted in Fig.1, for the cases when dark energy effects $y_2$ are taken into account, (Fig.1); when only GW effects to the orbit, $y_1$, are accounted for, but the new dark energy effects derived here are not included in calculating $R[t]$, (Fig.2).

ii) The second example is:  {\it a Radio Galaxy OJ287} observed in 2008 with {\it VLBA} \cite{vlba} which is suspected to be a  binary of two SBH's. This example is a bit trickier because one star is a lot heavier than the other and their total mass is $10^{10} M_{\ast}$. However, although algebraically messy, it is straightforward to estimate the effect of dark energy for this binary by using the expressions of Sec.2 for the case $m_1 \ne m_2$ and by replacing Eqn. \ref{masssimple} in order to derive Eqn. \ref{neworbit1}. The orbital separation is $ R_{0} \simeq 10^{20} m$ or equivalently $ 2 \beta \simeq 10^7$ at $t=0$. The current period is $T \simeq 12 yrs$,  
and their distance from us $r \simeq 3.5 Mly \simeq 10^{22}m$. If we were to approximate this binary with equal mass SBH's each with $a\simeq 10^{9}$ then we get $f_0 \simeq 10^{-9} Hz, h \simeq 10^{-20}$. We receive the emitted signal from the distance $r$ at a time $t_e \simeq 10^{16} s$ during which the orbit has changed by a factor $(R_0 y_1) = R_0 10^{-12} m, (R_0 y_2) = R_0 10^{-9} m$ due to Hulse-Taylor GW effect and the dark energy effect respectively. If dark energy is ignore ($y_2 = 0$), then merging occurs in about $t_{\ast} = 10^{12} sec \simeq 10,000 yrs$. Including the effect of dark energy accelerates the merging time by three orders of magnitude for the case $(1+w)>0$ or stops merging and rips the stars apart for the case of phantom energy $(1+w) < 0$.

As mentioned, binaries are observed with the {\it LIGO, LISA, VLBA, SDSS, Chandra} experiments. It is possible that more SBH's binaries will be found in the near future by these experiments. Among other things, the SBH's binary measurements will allow us to distinguish more accurately whether dark energy lies above or below the cc boundary $w = -1$ and to pin down the value of $w[z]$, by making use of the simple method proposed in this letter. 
However, the examples discussed here are sufficient to illustrate the power of our method for using SBH binaries to obtain information about dark energy and its equation of state $w[z]$ , even with existing GW  and binary data, while avoiding background noise issues, inheritant of the large scale structure in the universe.

\medskip

Acknowledgment:  L.M-H is supported in part by DOE grant DE-FG02-06ER1418, NSF grant PHY-0553312 and fqxi grant.

\pagebreak


\begin{thebibliography}{99}
\bibitem{wmap} E. Komatsu et al.,''Five-years Wilkinson microwave anisotropy probe (WMAP) observations:cosmological interpretation'',[astro-ph/0803.0547].

\bibitem{SDSS} By SDSS Collaboration (Jennifer K. Adelman-McCarthy et al.), Astrophys.J.Suppl.175:297-313, (2008), [astro-ph/0707.3413];
M.~E.~C.~Swanson, M.~Tegmark, M.~Blanton, I.~Zehavi, Mon.Not.Roy.Astron.Soc.385:1635-1655, (2008),[astro-ph/0702584]; R.~R.~Gibson, W.~N.~Brandt, D.~P.~Schneider, [astro-ph/0808.2603]
\bibitem{lauramelchiorri} A.~Melchiorri,
L. ~Mersini, C.~J.~Odman and M.~Trodden, Phys. \ Rev. {\bf D 68}, 043509 (2003).
\bibitem{accelerate} A. G. Reiss et al., Astroph. J {\bf 116}, 1009 (1998); 
S. Perlmutter et al., Astroph. J {\bf 517}, 565 (1998);  
P. H. Garnovich et al., ApJ {\bf 507}, 74 (1998). 
\bibitem{quintessence} .
Li-Min Wang, R.R. Caldwell, J.P. Ostriker, Paul J. Steinhardt, Astrophys.J. {\bf 530}:17-35,2000; ``An introduction to quintessence'', R.~R.~Caldwell, (2000), {\bf Braz.J.Phys.30}:215-229,2000.
\bibitem{kessence}
C.~Armendariz-Picon, V.~F.~Mukhanov, P.~J.~Steinhardt, Phys. \ Rev. \ Lett. {\bf 85}:4438-4441,(2000); C.~Armendariz-Picon, V.~F.~Mukhanov, P.~J.~Steinhardt,
Phys.Rev.D63:103510, (2001).

\bibitem{transplanck} L.~Mersini-Houghton, M.~Bastero-Gil, P.~Kanti, Phys. \ Rev. {\bf D64}:043508,2001, [hep-ph/0101210]; . 
M.~Bastero-Gil, L.~Mersini-Houghton, Phys. \ Rev. {\bf D65}:023502,(2002),[astro-ph/0107256] and, [hep-th/0212153]; 
M.~Bastero-Gil, L.~Mersini-Houghton, Phys. \ Rev. {\bf D67}:103519,(2003), [hep-th/0205271]; .
M.~Bastero-Gil, P.~H.~Frampton, L.~Mersini-Houghton, Phys. \ Rev. {\bf D65}:106002,(2002).  
\bibitem{eos} Y.~Wang, M.~Tegmark, Phys. \ Rev. \ Lett.{bf 92}:241302,2004; D.~Huterer and M.~S.~Turner, Phys. \ Rev. {\bf D 64}, 123527 ( 2001); 
E.~Linder, Phys.\ Rev. \ Lett.{bf 90}, 091301 (2003).

\bibitem{phantom} R.~R.~Caldwell, Phys. \ Lett. {\bf B545}, 23 (2002); R.~R.~Caldwell,M.~Kamionkowski and N.~N.~Weinberg, Phys. \ Rev. \ Lett. {\bf 91}, 071301 (2003
\bibitem{crunch} V.~Faraoni, W.~Israel, Phys. \ Rev. {\bf D71}:064017,2005, [gr-qc/0503005]; 
M.~Bouhmadi-Lopez, J.~Jimenez Madrid, {\bf JCAP 0505}:005, (2005),[astro-ph/0404540]; L.~Chimento, R.~Lazkoz, Mod. \ Phys. \ Lett. {\bf A19}:2479-2484,(2004), [gr-qc/0405020]. 

\bibitem{laurads} L. Mersini-Houghton, [arXiv:gr-qc/0609006]. 
\bibitem{eternalds} G.~W.~Gibbons and S.~W.~Hawking, Phys.~Rev.~ {\bf D15} 2738 (1977).

\bibitem{babichev} T.~Jacobson, Phys. \ Rev. \ Lett. {\bf 83}, 2699 (1999);
E.~Babichev, V.~Dokuchaev, Y.~Eroshenko, [gr-qc/0402089], Phys. \ Rev. \ Lett. {\bf 93}; P.~F.~Gonzalez-Diaz, [astro-ph/0312579];
E.~Babichev, V.~Dokuchaev, Y.~Eroshenko (Moscow, INR), J. \ Exp. \ Theor.\ Phys.{\bf 100:528-538} ,(2005), [astro-ph/0505618], and [gr-qc/0507119]. 
\bibitem{magueijo} R.~Bean and J.~Magueijo, Phys. \ Rev. {\bf D 66}, 063505 (2002)
\bibitem{thorn} {\it www.ligo-la.caltech.edu/contents/overviews.htm; elmer.tapir.caltech.edu/php237/}.
\bibitem{ostriker} J.~Hennawi and J.~Ostriker, [astro-ph/0108203].
\bibitem{bounce} M.~G.~Brown, K.~Freese, W.~H.~Kinney, {\bf JCAP 0803}:002,(2008), [astro-ph/0405353]; K.~Freese, M.~G.~Brown, W.~H.~Kinney, [astro-ph/0802.2583];  
M.~Bastero-Gil, K.~Freese, L.~ Mersini-Houghton (Syracuse U.), Phys. \ Rev. {\bf D68}:123514,(2003), [hep-ph/0306289];P.~Steinhardt and N.~Turok, Phys. \ Rev. {\bf D65}: 126003, (2002); J.~Khoury, P.~Steinhardt, and N.~Turok, Phys. \ Rev. \ Lett. {\bf 92}: 031302 (2004). 

\bibitem{ligo} LIGO Scientific Collaboration, "LIGO: The Laser Interferometer
Gravitational-Wave Observatory.", [gr-qc/0711.3041].

\bibitem{chandra} .
Elena Gallo,Jeroen Homan,Peter Jonker,John Tomsick,[astro-ph/0806.3491].
\bibitem{vlba}
G. B. Taylor, C. Rodriguez, R. T. Zavala, A. B. Peck, L. K. Pollack and R. W. Romani (2006). Imaging compact supermassive binary black holes with Very Long Baseline Interferometry. Proceedings of the International Astronomical Union, 2, pp 269-272.



\bibitem{fred} F. C. Adams and G. Laughlin, Rev. \ Mod. \ Phys. {\bf 69}, 337 (1997). 
\bibitem{fredlaura} L.~Mersini-Houghton and F.Adams, Class. \ Quant. \ Grav. (2008).







\end{thebibliography}
\end{document}